\documentclass[5p, preprint]{elsarticle}
\usepackage{amsmath}
\usepackage{txfonts}
\usepackage{graphicx}
\usepackage{dcolumn}
\usepackage{bm}
\usepackage{float}
\usepackage[section]{placeins}
\usepackage{float}
\usepackage{multirow}
\usepackage{comment}
\usepackage{color}
\usepackage{slashed}
\usepackage{ulem}
\usepackage{hyperref}
\usepackage{lineno}
\modulolinenumbers[2]

\newcommand{\delete}{\bgroup\markoverwith{\textcolor{red}{\rule[0.5ex]{2pt}{1pt}}}\ULon}

\begin{document}
\begin{frontmatter}
\title{Magicity of neutron-rich nuclei within relativistic self-consistent approaches}
\author[LZU,Lyon,Orsay]{Jia Jie Li\corref{cor}}
\ead{j.li@ipnl.in2p3.fr}
\cortext[cor]{Corresponding author.}
\author[Lyon]{J\'{e}r\^{o}me Margueron}
\author[LZU,Klab]{Wen Hui Long}
\author[Orsay]{Nguyen Van Giai}
\address[LZU]{School of Nuclear Science and Technology, Lanzhou University, Lanzhou 730000, China}
\address[Lyon]{Institut de Physique Nucl\'{e}aire de Lyon, IN2P3-CNRS, Universit\'{e} Claude Bernard Lyon 1, F-69622 Villeurbanne Cedex, France}
\address[Orsay]{Institut de Physique Nucl\'{e}aire, IN2P3-CNRS, Universit\'{e} Paris-Sud, F-91406 Orsay Cedex, France}
\address[Klab]{Key Laboratory of Special Function Materials and Structure Design, Ministry of Education, Lanzhou 730000, China}
\begin{abstract}
The formation of new shell gaps in intermediate mass neutron-rich nuclei is investigated within the relativistic Hartree-Fock-Bogoliubov theory, and the role of the Lorentz pseudo-vector and tensor interactions is analyzed. Based on the Foldy-Wouthuysen transformation, we discuss in detail the role played by the different terms of the Lorentz pseudo-vector and tensor interactions in the appearing of the $N=16$, 32 and 34 shell gaps. The nuclei $^{24}$O, $^{48}$Si and $^{52,54}$Ca are predicted with a large shell gap and zero ($^{24}$O, $^{52}$Ca) or almost zero ($^{48}$Si, $^{54}$Ca) pairing gap, making them candidates for new magic numbers in exotic nuclei. We find from our analysis that the Lorentz pseudo-vector and tensor interactions induce very specific evolutions of single-particle energies, which could clearly sign their presence and reveal the need for relativistic approaches with exchange interactions.
\end{abstract}
\end{frontmatter}
%
%
\section{Introduction}

In recent years the role of the tensor interaction in nuclei has gained renewed interest~\cite{Sagawa2014}.
Essentially induced by the Lorentz pseudo-vector (PV) pion and the Lorentz tensor (T) rho meson-nucleon couplings,
the tensor interaction is known to play an important role in the binding of light nuclei, such as the deuteron.
At this point, we must make clear the terminology employed throughout this paper.
The name "tensor interaction" will be used for the part of the non-relativistic nucleon-nucleon $NN$ interaction
which behaves as an irreducible second-rank tensor~\cite{Brink1994}.
Such a "tensor interaction" can originate from a non-relativistic reduction of PV or T meson-nucleon couplings.
Note, however, that the PV and T couplings contribute only to the exchange part of the nucleon-nucleon interaction, not to its direct part.
Therefore, relativistic approaches such as the relativistic mean field (RMF) where Fock terms are dropped will not lead to any tensor interaction.
It was however difficult to assign a clear effect of the tensor interaction in medium-heavy and heavy nuclei.
As a consequence, many-body approaches going from the non-relativistic Skyrme and Gogny effective interactions to the RMF approach, see e.g. Ref.~\cite{Bender2003} and references therein,
have simply ignored the presence of the tensor interaction, without loss of precision with respect to global properties of nuclei, e.g., masses or charge radii.
The situation has recently changed when it was shown that the nuclear tensor force could play a very important role in the isotopic evolution of some
single-particle (s.p.) states~\cite{Otsuka2005, Otsuka2006, Brown2006, Colo2007, Brink2007, Lesinski2007, Long2008, Nakada2008, Lalazissis2009, Otsuka2010, Smirnova2010, Anguiano2012},
causing the disappearance of the usual magic numbers and the emergence of new ones in some extreme cases of very neutron-rich nuclei~\cite{Sorlin2008, Nakada2010, Grasso2014, Yuksel2014}.
It is now clear that the modelling of exotic nuclei requires the development of more complete nuclear effective interactions that include the nuclear tensor force.

Employing the non-relativistic approaches based on Skyrme interactions, the improvement of the isotopic evolution of s.p. energies
by introducing the tensor force was found to be systematically correlated to a degradation of the binding energy~\cite{Lesinski2007}.
The tensor interaction considered in Refs.~\cite{Colo2007, Brink2007, Lesinski2007, Grasso2014, Yuksel2014} is a non-relativistic contact interaction.
The effects of a finite-range form of the tensor interaction were also explored in the case of the finite-range Gogny force~\cite{Otsuka2006, Anguiano2012}.
It was shown that the finite-range tensor force has a large impact on s.p. energies along isotopic and isotonic chains but its effect on binding energies was not discussed.
Finally, we must mention the finite-range M3Y interactions which contain the tensor interaction and are satisfactory for binding energies and s.p. spectra~\cite{Nakada2008, Nakada2010}.
In this approach the in-medium tensor interaction is not much modified and resembles to a large extent that of the original one in the bare $NN$ interaction.
Shell model calculations with $V_{\textsl{low}\,k}$ low-momentum interactions have also been performed, including systematic comparisons with and without tensor interactions,
illustrating the important role of the tensor terms~\cite{Otsuka2010}.

In all the previously mentioned studies, the tensor interaction at work was mostly explored in a non-relativistic framework.
In a relativistic framework, it however contributes only to the Fock diagrams,
where not only the $\pi$-$N$ and Lorentz tensor $\rho$-$N$ couplings play an important role but the other meson-nucleon couplings also carry considerable tensor force components~\cite{Jiang2015}.
Since in the RMF approaches~\cite{Vretenara2005, Meng2006} the Fock diagrams are simply dropped, the relativistic Hartree-Fock (RHF) approach~\cite{Lalazissis2009, Long2006, Bouyssy1987}
becomes the only relativistic model which generates a tensor force.
In this work, we compare the predictions based on relativistic Hartree-Fock-Bogoliubov (RHFB)~\cite{Long2010} and relativistic Hartree-Bogoliubov (RHB)~\cite{Vretenara2005, Meng2006} approaches,
which can both reproduce satisfactorily global properties such as binding energies and radii of finite nuclei.
In the RHFB approach, the $\pi$-pseudo-vector ($\pi$-PV) and $\rho$-tensor ($\rho$-T) couplings can be taken into account while they do not contribute in the RHB models.
Considering the Foldy-Wouthuysen (FW) transformation~\cite{Foldy1950}, the relation between the Lorentz type and rank-2 tensor forces could be analyzed,
as well as their respective contributions to s.p. energies and shell evolutions.
We investigate the shell evolution of Ca isotopes, $N=16$, 32 and 34 isotones and analyze the appearance of a large shell gap in $^{24}$O and $^{52,54}$Ca
making $N=16$, 32 and 34 candidates for being new magic numbers in neutron-rich nuclei.

This paper is organized in the following way: In Sec.~\ref{Formalisms} the relativistic framework is briefly recalled, as well as the non-relativistic reduction of the Lorentz PV and T couplings.
The results and discussions for the selected isotopes and isotones are given in Sec.~\ref{Results}, and the summary and conclusions are presented in Sec.~\ref{Summary}.

\section{Lorentz pseudo-vector and tensor interactions in a relativistic Hartree-Fock model}\label{Formalisms}

Within the relativistic framework, Lorentz PV $\pi$-$N$ coupling and T $\rho$-$N$ coupling
can be introduced into the effective Lagrangian. The corresponding interaction
vertices are:
\begin{subequations}
\begin{align}
\Gamma^{PV}_\pi(1,2) &\equiv -\frac{1}{m^2_\pi} \Big(f_\pi\vec{\tau}\gamma_5\gamma_{\mu}\partial^{\mu}\Big)_1\cdot \Big(f_\pi\vec{\tau}\gamma_5\gamma_{\nu} \partial^{\nu}\Big)_2, \\
\Gamma^{T}_\rho(1,2) &\equiv +\frac{1}{4M^2} \Big(f_\rho\vec{\tau}\sigma_{\lambda \mu}\partial^\mu\Big)_1\cdot \Big(f_\rho\vec{\tau}\sigma^{\lambda \nu} \partial_\nu\Big)_2,
\end{align}
\end{subequations}
where $m_\pi$ and $M$ denote the mass of the $\pi$ meson and of the nucleon, respectively, $f_\pi$ and $f_\rho$ are the coupling constants of the $\pi$-PV and $\rho$-T meson-nucleon vertices.
The other coupling channels, namely the isoscalar scalar $\sigma$ ($\sigma$-S) and vector $\omega$ ($\omega$-V), the isovector vector $\rho$ ($\rho$-V),
and the vector photon $A$ ($A$-V) couplings are identical to those presented in Refs.~\cite{Bouyssy1987, Long2007}.

In the non-relativistic limit, using the FW transformation~\cite{Foldy1950}, the finite-range part of the one-$\pi$ exchange potential $V_\pi$ can be divided into two terms:
the tensor and the central potentials, $V^T_\pi$ and $V_\pi^C$,
\begin{subequations}
\begin{align}
V^T_\pi(\bm{r}) &= \frac{1}{3} f^2_\pi \,\bm{\tau}_1 \cdot \bm{\tau}_2 \, S_{12} \, \Bigg(1 + \frac{3}{m_\pi r} + \frac{3}{(m_\pi r)^2}\Bigg)\frac{e^{-m_\pi r }}{r},\\
V^C_\pi(\bm{r}) &= \frac{1}{3} f^2_\pi \,\bm{\tau}_1 \cdot \bm{\tau}_2 \,\,\bm{\sigma}_1 \cdot \bm{\sigma}_2 \,\frac{e^{-m_\pi r }}{r},
\end{align}
\end{subequations}
where $S_{12}$ is a standard rank-2 tensor operator,
\begin{align}
S_{12}(\bm{r}) \equiv 3(\bm{\sigma}_1\cdot\bm{e}_r)(\bm{\sigma}_2\cdot\bm{e}_r)-\bm{\sigma}_1\cdot\bm{\sigma}_2.
\end{align}

The FW transformation~\cite{Foldy1950} could similarly be applied to the $V_\rho$ potential originating from the Lorentz T $\rho$-$N$ coupling. One must notice that,
in the non-relativistic formalism, the central and rank-2 tensor forces generally appear independently from each other, while here they originate from the same interaction vertices.
The central term $V^C_{\pi (\rho)}$ plays an important role in determining the shell structure, which is different but as important as the rank-2 tensor term $V^T_{\pi (\rho)}$~\cite{Nakada2008, Otsuka2001}.

\section{Results and discussions}\label{Results}

In our relativistic approaches the Lorentz PV and T couplings, when present, cannot be switched off without degrading the accuracy of the model.
We instead compare various Lagrangians having, or not, the Lorentz PV and T terms as explained below.
For the RHFB approach, we consider the effective interactions PKA1~\cite{Long2007} and PKO3~\cite{Long2008}.
The former is the most complete relativistic model to date, taking both $\pi$-PV and $\rho$-T into account, while the latter does not contain the $\rho$-T coupling.
All meson-nucleon couplings are density-dependent and their values have been adjusted in previous studies.
Two density-dependent RHB models are also used throughout this work: DD-ME2~\cite{Lalazissis2005} and DD-ME$\delta$~\cite{Roca2011}.
The DD-ME$\delta$ differs from DD-ME2 by the inclusion of the $\delta$-meson, which leads to different proton ($\pi$) and neutron ($\nu$) Dirac masses.
Since all these modelings do not contain originally a term describing the pairing effects, the Gogny force D1S~\cite{Berger1984} is employed in the pairing channel.
DD-ME2 and DD-ME$\delta$ do not have tensor terms (neither of Lorentz type, or of rank-2 type).
Since these four effective Lagrangians reproduce equally well the energies and radii of finite nuclei,
we could compare in the following the impact of the tensor force on more detailed quantities such as s.p. energies,
within a spherical model which was described in Refs.~\cite{Long2010, Li2014}.

\subsection{Lorentz pseudo-vector and tensor effects on Ca isotopes}

The analysis of the oxygen isotopes has revealed the very important role of the spin-isospin properties of the nuclear interaction and their impact on the magicity of $^{24}$O~\cite{Nakada2008, Otsuka2001}.
The origin of this strong spin-isospin interaction is directly related, in a one boson-exchange picture, to the contribution of $\pi$ and tensor-$\rho$ mesons.
In a similar way, it is interesting to explore the role of these mesons along the next semi-magic $Z=20$ (resp. $N=20$) isotopic (resp. isotonic) chains with respect to the neutron (resp. proton) shell evolution. Therefore, it is expected that, since there exist a large number of measured properties, the analysis of the theoretical predictions could shed light on the role of these meson fields, and more precisely, of the tensor force. Notice that $^{40}$Ca is spin-saturated in the proton and neutron shells, and the tensor interaction that contributes in calcium isotopes above $^{40}$Ca is mostly active between a few neutron states: mostly $\nu2p$ and $\nu1f$ states. Calcium isotopes therefore provide an ideal isotopic chain for the theoretical and experimental analysis of the neutron-neutron tensor interaction and for studying its role in the formation and evolution of neutron shells in medium mass nuclei.

\begin{figure}[tb]
\centering
\includegraphics[width = 0.40\textwidth]{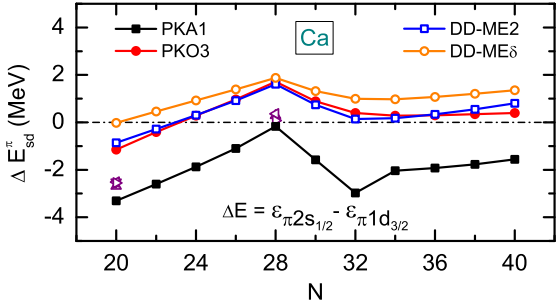}
\caption{(Color online.) Energy difference $\Delta E^\pi_{sd} = \varepsilon_{\pi2s_{1/2}}-\varepsilon_{\pi1d_{3/2}}$ of Ca isotopes. The experimental values of $^{40}$Ca and $^{48}$Ca are taken from Ref.~\cite{Grawe2007} (up-triangle), Ref.~\cite{Schwierz2007} (down-triangle), Ref.~\cite{Isakov2002} (left-triangle), and Ref.~\cite{Oros1996} (right-triangle), respectively.
}
\label{fig:PSD}
\end{figure}

We first analyze the question of the possible proton level inversion in $^{48}$Ca. The normal level sequence being $\{\pi 2s_{1/2}, \pi 1d_{3/2}\}$,
it has been suggested that the inversion of this sequence if well marked, could induce the formation of a proton depletion, a kind of bubble structure,
in the neighboring $^{46}$Ar, reducing the spin-orbit (SO) splitting~\cite{Todd2004}. The energy difference $\Delta E^\pi_{sd} = \varepsilon_{\pi2s_{1/2}} - \varepsilon_{\pi1d_{3/2}}$
along the calcium isotopic chain is shown in Fig.~\ref{fig:PSD} for the selected effective Lagrangians: PKA1, PKO3, DD-ME2 and DD-ME$\delta$.
The experimental data of the relevant s.p. energies in both $^{40}$Ca and $^{48}$Ca are also reported in Refs.~\cite{Grawe2007, Schwierz2007, Isakov2002, Oros1996}.
Experimentally, s.p. energies are difficult to determine due to the strength fragmentation caused by the particle-vibration coupling~\cite{Bernard1980, Litvinova2011, Baldo2015}.
Among these effective Lagrangians, it appears that only the RHFB-PKA1 model gives a satisfactory agreement with the experimental results and predicts an almost perfect degeneracy between
the $\pi 2s_{1/2}$ and $\pi 1d_{3/2}$ states, consistent with the data (see Fig.~\ref{fig:PSD}).
The other effective Lagrangians, which do not contain the Lorentz PV and T couplings or only partially (such as PKO3 which has the PV coupling),
fail to reproduce the experimental points, and a level inversion is even predicted around $^{48}$Ca, in contradiction with the data.
From the comparison shown in Fig.~\ref{fig:PSD}, we deduce that the tensor-$\rho$ meson-nucleon coupling, which can be treated as a mixture of central and tensor forces,
is an important ingredient in order to reproduce the s.p. spectra in Ca isotopes. The importance of the tensor force in this context has also been stressed by non-relativistic approaches~\cite{Smirnova2010, Grasso2007, Nakada2013}. It is worthwhile to notice that the two proton states $\pi 2s_{1/2}$ and $\pi 1d_{3/2}$ are pseudo-spin (PS) partners~\cite{Ginocchio2005, Liang2015}.
The energy difference $\Delta E^\pi_{sd}$ could therefore be interpreted as a measure of the PS degeneracy.
In this case, the Lorentz tensor $\rho$-field is also an important ingredient for discussing the occurrence of PS degeneracy in $^{48}$Ca.

We now turn to the discussion of the emergence of new shell closures in exotic calcium isotopes which have been indicated by recent experiments.
First, the high excitation energy of $^{52}$Ca, compared to those of neighboring nuclei, favors a possible new $N=32$ shell closure~\cite{Gade2006}.
Then, a confirmation was obtained from recent high-precision mass measurements of several isotopes ranging from $^{51}$Ca up to $^{54}$Ca~\cite{Wienholtz2013}.
In neighboring titanium and chromium nuclei, a high excitation energy was also observed indicating the robustness of the $N = 32$ magic number in this region of the nuclear chart~\cite{Prisciandaro2001, Dinca2005}.
Additionally, from the very recent measurement of the $2^+_1$ energy in $^{54}$Ca which was found to be only $\sim$500~keV below that in $^{52}$Ca~\cite{Steppenbeck2013},
it is also possible that $^{54}$Ca could be magic as well. If confirmed it would be the first example of two successive magic numbers ($N=32$ and 34) in the same element.
It should however be noticed that the possible $N=34$ magic number is less marked in the neighboring nuclei such as titanium where the $2^+_1$ excitation energy
was found smaller than in calcium~\cite{Steppenbeck2013, Liddick2004}.

\begin{figure}[tb]
\centering
\includegraphics[width = 0.48\textwidth]{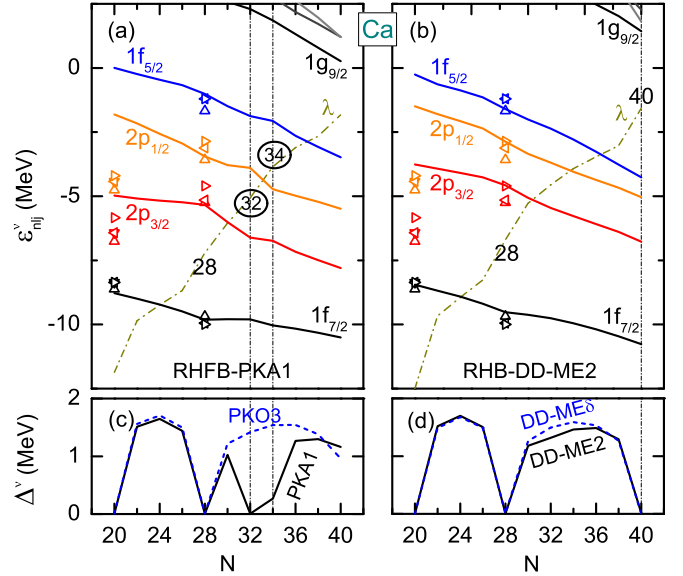}
\caption{(Color online.)
Upper panels: Neutron s.p. spectra of Ca isotopes, extracted from the RHFB calculations with PKA1 (plot a) and the RHB ones with DD-ME2 (plot b). Lower panels: Neutron pairing gaps determined by the RHFB (plot c) and RHB (plot d) calculations.
The experimental data (in various triangles) are taken from Refs.~\cite{Grawe2007, Schwierz2007, Isakov2002, Oros1996}, see Fig.~\ref{fig:PSD} for the convention of symbols.
}
\label{fig:CA}
\end{figure}

The role of the different components of the nuclear interaction can be analyzed from various perspectives.
The mean-field in non-relativistic Skyrme approaches allows to analyze the respective roles of the central, SO and effective mass terms~\cite{Grasso2007};
the tensor force in M3Y-type models was found to be important in reproducing the experimental trends~\cite{Nakada2010};
the comparison between the relativistic and non-relativistic approaches shows that $N$ or $Z=8$, 20 are suitable for fitting the tensor parameters~\cite{Moreno2010},
and finally, calcium isotopes have been used as benchmark nuclei to illustrate the role of the three-body (TB) force~\cite{Hagen2012, Soma2014, Hergert2014}.
Concerning the last study, it is worth noticing a recent work showing that a properly optimized chiral two-body interaction can also describe many aspects of nuclear structure in oxygen and calcium isotopes without explicitly invoking TB force~\cite{Ekstrom2013}.
Just after the measurements of neutron-rich calcium isotopes~\cite{Wienholtz2013, Steppenbeck2013}, it was argued from a Skyrme Hartree-Fock (SHF) approach that the magicity of $^{52}$Ca and $^{54}$Ca might be reproduced selecting fitted tensor parameters~\cite{Grasso2014}.
Pairing correlations have however not been considered and the magicity has been inferred only from the size of the s.p. gaps.
It is known that relatively large s.p. gaps could still be overcome by pairing correlations leading to fractional occupations above the expected s.p. shell gap.
As proposed in a non-relativistic approach based on the Skyrme force~\cite{Yuksel2014}, a criterion for the magicity should thus also refer to the occupation number above the s.p. gap which,
in case of magicity, should be found to be zero.

Fig.~\ref{fig:CA} shows the isotopic evolution of the canonical s.p. energies (upper panels) and average pairing gaps (lower panels) from $^{40}$Ca to $^{60}$Ca, 
going over the possible new magic numbers at $N=32$ and 34, and using RH(F)B approach with the selected effective Lagrangians. In Figs.~\ref{fig:CA} (a) and (b) there are shown the evolutions of the neutron s.p. energies around the chemical potential determined by RH(F)B calculations with PKA1 and DD-ME2, respectively.
We remind that the RHB-DD-ME2 model contains neither the Lorentz tensor nor rank-2 tensor force components.
In Fig.~\ref{fig:CA}, the experimental values taken from Refs.~\cite{Grawe2007, Schwierz2007, Isakov2002, Oros1996} are also shown.
It should be noticed that we only take the experimental data as a reference due to the limit of mean-field approach.
In the present calculations, the correlations beyond mean field~\cite{Bernard1980, Litvinova2011, Baldo2015, Duguet2012, Duguet2015}, such as the particle-vibration couplings, are not taken into account.
Another fact is that among the selected s.p. states some states such as the $\nu2p_{1/2}$ in $^{40}$Ca or $\nu1f_{5/2}$ in $^{48}$Ca are highly fragmented.

The general feature reflected in Fig.~\ref{fig:CA} is that, as the number of neutrons increases, the s.p. energies decrease due to the enhanced mean-field potential.
This trend is smooth for the RHB-DD-ME2, as well as for the other non-presented models RHFB-PKO3 and RHB-DD-ME$\delta$, while we observe abrupt changes in the RHFB-PKA1 results when the s.p. energies cross the chemical potential. Notice moreover that $^{40}$Ca is spin-saturated, while the spin asymmetry reaches a maximum at $^{52}$Ca.
We have analyzed the isotopic evolution of the two-body interaction matrix elements $V_{ii^\prime}$ between neutron valence orbits and we found that the couplings are quite constant for the RHB-DD-ME2 while for the RHFB-PKA1 they are slightly more dependent on the number of neutrons.
In particular, it is observed a reduction of the coupling between $\nu 2p_{1/2}$ and $\nu 2p_{3/2}$ states at $N=32$ by about 40\% as compared to the values at $N=30$ and 34.
This reduction is associated with the emergence of a shell gap at $N=32$. We also observe that the coupling between $\nu 2p_{1/2}$ and $\nu 2p_{3/2}$ states is reduced by a factor 2 with the RHFB-PKA1 as compared to the RHB-DD-ME2, which explains why the gap at $N=32$ is not so distinct for the RHB-DD-ME2.
In contrast, although both Lagrangians give almost identical coupling strengths between the $\nu 2p_{1/2}$ and  $\nu 1f_{5/2}$ states, only the RHFB-PKA1 shows a gap at $N=34$.
The gaps at $N=32$ and 34 are therefore related to different parts of the nuclear interaction, and we will analyze it with respect to the Lorentz PV and T interactions in the following.

\begin{table}[tb]
\caption{Energy difference $\Delta E(i,i^\prime) \equiv \varepsilon_i - \varepsilon_{i^\prime}$ (in MeV) in Ni, Ca and Si at $N=32$ and 34. The results correspond to RHFB-PKA1, PKO3 and RHB-DD-ME2 Lagrangians. See text for details.}
\setlength{\tabcolsep}{6.8pt}
\label{tab:ED}
\begin{tabular}{cccrcc}
\hline
Force                & $\Delta E(i,i^\prime)$    &  $N$ &   Ni   &     Ca      &   Si        \\
\hline
\multirow{2}*{PKA1}  &($\nu2p_{1/2},\nu2p_{3/2}$)&  32  &   1.51 &\textbf{2.72}&  0.81       \\
                     &($\nu1f_{5/2},\nu2p_{1/2}$)&  34  &   1.04 &\textbf{2.45}&\textbf{4.05}\\
\hline
\multirow{2}*{PKO3}  &($\nu2p_{1/2},\nu2p_{3/2}$)&  32  &   1.22 &    1.69     &  0.68       \\
                     &($\nu1f_{5/2},\nu2p_{1/2}$)&  34  &$-$1.72 &    0.77     &  2.72       \\
\hline
\multirow{2}*{DD-ME2}&($\nu2p_{1/2},\nu2p_{3/2}$)&  32  &   1.58 &    1.76     &  0.92       \\
                     &($\nu1f_{5/2},\nu2p_{1/2}$)&  34  &$-$1.23 &    1.21     &  3.18       \\
\hline
\end{tabular}
\end{table}

As discussed before, the magicity is not solely related to an increase of the shell gap, but also to a quenching of the pairing correlations at magic numbers.
As seen from Figs.~\ref{fig:CA}(c) and \ref{fig:CA}(d), the suppression of the neutron pairing gaps at traditional magic numbers $N=20$ and 28 is confirmed by all the models considered here.
In the RHFB-PKA1 results, there is an additional suppression of the neutron pairing gap at $N=32$ which is not predicted by the other Lagrangians.
This large quenching for $N=32$ is an additional hint which suggests that the $N=32$ magic number should be analyzed in the light of the Lorentz PV and T interactions.
A weaker, but still distinct and important, quenching is also predicted by the RHFB-PKA1 at $N=34$, which may suggest as well $N=34$ to be a submagic number.
It is also interesting to notice the predictions at $N=40$: the RHB models indicate a quenching of the pairing correlations at $N=40$, while the RHFB ones predict a small decrease but not a quenching.
The origin of these discrepancies is related to the Fock term and to the effect of the Lorentz PV and T interactions.
The most advanced Lagrangians presented here (PKA1 and PKO3) therefore do not predict $N=40$ as a magic number for $^{60}$Ca.

\subsection{Evolution along $N=32, 34$ down to the drip line}
\begin{figure}[tb]
\centering
\includegraphics[width = 0.48\textwidth]{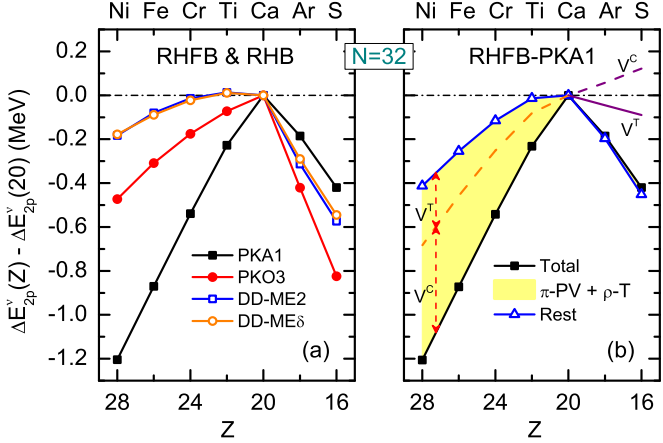}
\caption{(Color online.) (a) The SO splitting $\Delta E^\nu_{2p}$ of $\nu2p$ along the $N = 32$ isotonic chain, calculated with the RHFB (full symbols) and RHB (open symbols). (b) Detailed contributions to the SO splitting from the $\pi$-PV and $\rho$-T couplings, in comparison with those from the other channels. The results are obtained with the RHFB-PKA1 Lagrangian.}
\label{fig:D2P}
\end{figure}

We now analyze the contribution of the neutron-proton interaction to the SO splittings of $\nu2p$ states.
To do so, we fix the neutron number as $N=32$ and vary the proton number from $^{60}$Ni down to $^{52}$Ca, 
and along the isotonic line the valence protons are gradually removed from the $\pi1f_{7/2}$ state (which is a $j_>$ state). 
The energy differences associated with the $N=32$ and $34$ shell gaps are given in Table~\ref{tab:ED} for a set of selected isotones going from the stability line down to the drip line: $^{60,62}$Ni, $^{52,54}$Ca and $^{46,48}$Si. Comparing the $\nu2p$ SO splittings ($N=32$) given by different effective Lagrangians, it is interesting to notice that the values are almost the same for $^{60}$Ni ($1.4\pm0.2$ MeV) and for $^{46}$Si ($0.8\pm0.1$ MeV), while they are quite different for $^{52}$Ca. A more systematic calculation along the $N=32$ isotonic line is shown in Fig.~\ref{fig:D2P} where Fig.~\ref{fig:D2P}(a) shows the comparison of the SO splittings between the models (PKA1, PKO3, DD-ME2 and DD-ME$\delta$) and Fig.~\ref{fig:D2P}(b) presents the detailed contributions of PKA1. To better analyze the isotonic evolution, we present the results with respect to the values calculated in $^{52}$Ca.

As shown in Fig.~\ref{fig:D2P}(a), the RHFB-PKA1 results show a distinct enhancement of the SO splitting from Ni to Ca ($\sim$1.2~MeV), 
while it is less pronounced for RHFB-PKO3 ($\sim$0.5~MeV) and much less for RHB-DD-ME2 and RHB-DD-ME$\delta$ ($\sim$0.2~MeV). 
On the other side, from S to Ca the situation is changed. Towards Ca, the RHFB-PKO3 results present a 0.8~MeV 
enhancement of the $\nu2p$ SO splitting ($\sim$0.5 and 0.4~MeV respectively for RHB and RHFB-PKA1).
Here, we truncate at S to make the figures more readable, but the conclusions extend beyond. 
To further analyze the origin of this behavior and the role of the Lorentz PV and T interactions, we have plotted in Fig.~\ref{fig:D2P}(b) the contribution from the Lorentz PV and T couplings in RHFB-PKA1,
which are decomposed into the rank-2 tensor term $V^T$ and the central term $V^C$, as compared to the other terms (Rest) associated with the kinetic energy term,
$\sigma$-S, $\omega$-V and $\rho$-V couplings. We find that the Lorentz PV and T couplings, namely the $\pi$-PV and $\rho$-T ones in PKA1, 
play a dominant role in determining the enhancement of the $\nu2p$ SO splitting from Ni to Ca, while from Ca to S, 
the contributions of the Lorentz PV and T couplings are close to zero. Such difference is due to the fact that the contributions of the central and rank-2 tensor terms,
namely $V^C$ and $V^T$, add up between Ni and Ca, while they mutually compensate between Ca and S. In addition, it is worth noticing the large contribution of the central
term to the SO splitting between Ni and Ca (70\% of the increase), while the rank-2 tensor accounts for only 30\% of the increase. 
In conclusion, it is illustrated in the case of the $N=32$ isotones that the effects of the Lorentz PV and T on the s.p. energy gaps are certainly not reducible to the contribution of the rank-2 tensor because the central term plays a more important role.

\begin{figure}[tb]
\centering
\includegraphics[width = 0.48\textwidth]{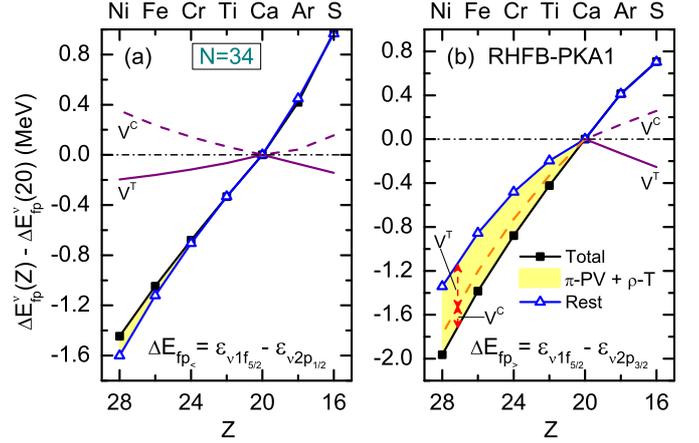}
\caption{(Color online.) Detailed contributions from the $\pi$-PV and $\rho$-T couplings to the energy differences
(a) $\Delta E^\nu_{fp_<}=\varepsilon_{\nu1f_{5/2}}-\varepsilon_{\nu2p_{1/2}}$ and
(b) $\Delta E^\nu_{fp_>}=\varepsilon_{\nu1f_{5/2}}-\varepsilon_{\nu2p_{3/2}}$ along the $N=34$ isotonic chain. The results correspond to the RHFB-PKA1 Lagrangian.}
\label{fig:DPF}
\end{figure}

We now turn to the $N=34$ gap which is appearing between the $\nu 2p_{1/2}$ and $\nu 1f_{5/2}$ states. Similarly to the $N=32$ case, we first refer to Table~\ref{tab:ED} where
the energy differences responsible for the $N=34$ shell gap, $\Delta E^\nu_{fp_<}=\varepsilon_{\nu1f_{5/2}}-\varepsilon_{\nu 2p_{1/2}}$, are given for Ni, Ca and Si. 
Only the RHFB-PKA1 model presents fairly distinct shell gap at $N=34$ for $^{54}$Ca and all the models predict much enhanced gap for $^{48}$Si. For $^{62}$Ni, 
the order of the states is even changed in the RHFB-PKO3 and RHB-DD-ME2 results. In order to clarify the role of the Lorentz PV and T in the $N=34$ case, 
we represent in Fig.~\ref{fig:DPF}(a) the evolution of the energy difference $\Delta E^\nu_{fp_<}$ shifted by that in $^{54}$Ca, while the energy difference involving the SO partner of the occupied state,
$\Delta E^\nu_{fp_>}=\varepsilon_{\nu1f_{5/2}}-\varepsilon_{\nu2p_{3/2}}$, is represented in Fig.~\ref{fig:DPF}(b). The role of the Lorentz PV and T is not so straightforward since the states are not SO partners. 
It is found, for PKA1, that the Lorentz PV and T forces present tiny contributions to the splitting between the $\nu2p_{1/2}$ and $\nu1f_{5/2}$ states. As shown in Fig.~\ref{fig:DPF}(a) the origin of this weakening is due to the near cancellation between the central and the rank-2 tensor components of the Lorentz PV and T couplings. This cancellation occurs for all isotones considered here, from Ni to S.

It is interesting to notice that the states $\{\nu 2p_{3/2}, \nu1f_{5/2}\}$, involving the SO partner of the occupied state of the $N=34$ gap, are PS partners and usually named $\nu 1\tilde{d}$, cf. Fig.~\ref{fig:CA}. The contribution to the PS splitting of Lorentz PV and T terms ($\pi$-PV and $\rho$-T) as well as that coming from the other terms (Rest) calculated with the RHFB-PKA1 model are represented in Fig.~\ref{fig:DPF}(b). It is observed that the other terms (Rest) largely contribute to the energy difference $\Delta E^\nu_{fp_>}$, while the Lorentz PV and T contribute only to about 30\% of the splitting. For the $N=34$ isotones, we observe that the $\nu 2p$ SO splitting is changing rather weakly from $Z=28$ to 20, and therefore the opening of the $N=34$ gap for $Z=20$ can be understood, to a large extent, from the evolution of the PS splitting. The shell gap at $N = 34$ can therefore be interpreted as a manifestation of a strong isospin-dependent PS splitting in which the Lorentz PV and T terms have only a weak impact.

It is worth comparing our results to very recent experimental analyses of shell gaps in $^{50}$Ar, $^{52,54}$Ca, $^{54}$Ti~\cite{Steppenbeck2015}.
Using RHFB-PKA1, we observe as well that the $N=32$ subshell gaps in $^{50}$Ar, $^{52}$Ca, $^{54}$Ti are similar in magnitude, see Fig.~\ref{fig:D2P}(a),
and that the $N=34$ subshell closure in $^{52}$Ar is larger than in $^{54}$Ca, see Fig.~\ref{fig:DPF}(a).
In addition, the shell gaps continue to increase between $^{52}$Ar and $^{50}$S, see Fig.~\ref{fig:DPF}(a),
and this tendency is confirmed for the next $N=34$ nucleus which is $^{48}$Si.
RHFB-PKA1 predicts a large gap for the drip line nucleus $^{48}$Si ($\sim$4.0~MeV) and a small pairing gap, while the other Lagrangian predict a small gap, see Table~\ref{tab:ED}.

In conclusion of this part, the effect of the Lorentz PV and T couplings ($\pi$-PV and $\rho$-T) on the $N=32$ shell gap is shown to be dominant at variance with the $N=34$ one. The origin of these shell gaps are indeed different: the $N=32$ shell gap is related to the $\nu 2p$ SO splitting while the $N=34$ shell gap can be related to the $\nu 1\tilde{d}$ PS splitting. The prediction of two successive magic numbers in Ca isotopes ($^{52,54}$Ca) is not related to the same origin.

\subsection{Magicity of $N=16$ in neutron-rich isotopes}

Based on the evolution of neutron separation energies and cross sections in light nuclei, it was originally proposed $N=16$ to be a new magic number lying between the usual $N=8$ and 20 for N, O and F nuclei~\cite{Ozawa2000}. In these nuclei, the s.p. gap at $N=16$ is occurring between the $\nu1d_{3/2}$ and $\nu2s_{1/2}$ states. Several experimental analyses have concluded as well that $N=16$ could be a magic number. For instance, from proton knockout reactions using $^{26}$F beam, the s.p. gap between these states was deduced to be 4.86(13)~MeV~\cite{Hoffman2008}. Another confirmation came also from the measurement of a relatively high excitation energy and a small $\beta_2$ value in $^{24}$O~\cite{Tshoo2012}. These results also indicate that $^{24}$O has a spherical character.
\begin{figure}[tb]
\centering
\includegraphics[width = 0.48\textwidth]{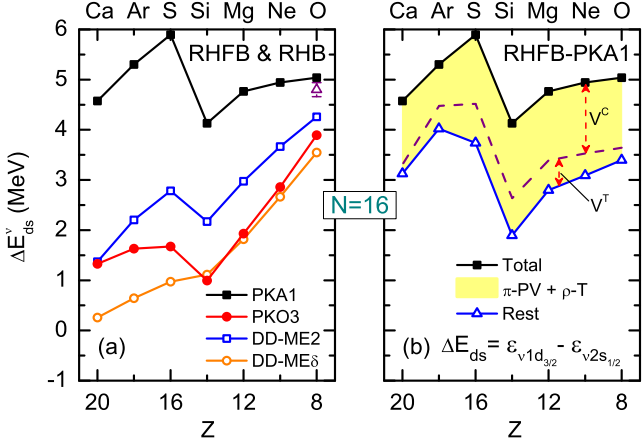}
\caption{(Color online.) (a) Energy difference $\Delta E^\nu_{ds} = \varepsilon_{\nu1d_{3/2}}-\varepsilon_{\nu2s_{1/2}}$ along the $N = 16$ isotonic chain, calculated with the RHFB (full symbols) and RHB (open symbols). The experimental value (triangle) of $^{24}$O~\cite{Hoffman2008} is also displayed as a reference. (b) Detailed contributions to $\Delta E^\nu_{ds}$ from the $\pi$-PV and $\rho$-T couplings, in comparison with those from the other channels. The results are obtained with the RHFB-PKA1 Lagrangian.}
\label{fig:NSD}
\end{figure}

Our predictions are shown in Fig.~\ref{fig:NSD} where we represent the energy difference between the $\nu 1d_{3/2}$ and $\nu2s_{1/2}$ states, namely $\Delta E^\nu_{ds} = \varepsilon_{\nu 1d_{3/2}}-\varepsilon_{\nu2s_{1/2}}$, along the $N=16$ isotonic chain from $Z=20$ down to 8. Here, we have used as pairing interaction the original Gogny force D1S modified by an adjusted strength factor $f = 1.10$, so as to reproduce the odd-even mass differences of oxygen isotopes~\cite{Wang2013}.
In Fig.~\ref{fig:NSD}(a) we compare the predictions of different effective Lagrangians, PKA1, PKO3, DD-ME2 and DD-ME$\delta$, with the experimental data in $^{24}$O~\cite{Hoffman2008}. The effective Lagrangian PKA1 predicts a large s.p. gap from $Z=8$ up to 20, as well as a s.p. gap in $^{24}$O close to the experimental value. At variance with this prediction, the three other effective Lagrangians (PKO3, DD-ME2, and DD-ME$\delta$) predict a strong reduction of the s.p. gap when going from $Z=8$ to 20, and the s.p. gap predicted for $^{24}$O is well below the data. In order to analyze the role played by the Lorentz PV and T couplings on the s.p. gap evolution, we present in Fig.~\ref{fig:NSD}(b) the contributions from the central and rank-2 tensor terms of the Lorentz PV and T couplings, as well as the other terms (Rest) for the RHFB-PKA1 model. For $Z=20$ down to 8, the Lorentz PV and T couplings contribute about 30-50\% of the s.p. gap, and it is interesting to note that the contribution from the central terms of the Lorentz PV and T couplings are dominant and the effect of the rank-2 tensor coupling is quite weak. The latter observation contradicts previous studies based on non-relativistic approaches using Skyrme and Gogny forces where a stronger rank-2 tensor force is necessary to reproduce s.p. evolutions~\cite{Anguiano2012}.
The contribution of the Lorentz PV and T tensor couplings (yellow region in Fig.~\ref{fig:NSD}) for $N=16$ is rather constant along the isotonic chain, and the observed $Z$-evolution is mostly related to the "Rest" terms and seems to be quite impacted by shell effects. A general feature observed with our relativistic models is that the shell gap is decreasing from O to Si, despite an increase around $Z=14$.

\section{Summary and conclusions}\label{Summary}

In summary, the formation of new shell gaps in neutron-rich nuclei is investigated within the RHFB theory and the role of the Lorentz PV and T interactions is studied in detail by comparing different Lagrangians with or without such terms. Based on the most complete RHFB-PKA1 effective Lagrangian, we confirm and predict that $^{48}$Si, $^{52,54}$Ca and $^{24}$O are the magic neutron-rich nuclei. In the case of $^{52}$Ca, the role of the Lorentz PV and T components is determinant, while it is less important for $^{24}$O, and negligible for $^{54}$Ca and $^{48}$Si. Analyzing shell evolutions along the $N=32$, 34 and $N=16$ isotonic chains, we observe that the global variation of the s.p. energies is due to the isoscalar component of the effective Lagrangian, while the more specific evolution of SO and PS partners is related to the Lorentz PV and T couplings ($\pi$-PV and $\rho$-T). Based on the FW transformation, we analyze the role played by the central and rank-2 tensor terms of the Lorentz PV and T couplings in the formation of the $N = 16, 32$ and 34 shell gaps. It is shown that those terms drive the distinct enhancement of the $N = 32$ gap along isotonic chains going from $^{60}$Ni to $^{52}$Ca, while the increase of the $N = 34$ gap from $^{62}$Ni to $^{54}$Ca, and $^{48}$Si, is mainly due to the other channels. The shell gap at $N = 34$ can be also interpreted as a manifestation of a strong isospin-dependent PS splitting. Finally, we also observe that the Lorentz PV and T couplings are not likely to support the appearance of $N = 40$ magicity in $^{60}$Ca.

In this study, we have illustrated in intermediate mass nuclei the very important role played by the Lorentz PV and T interactions which cannot be simply reduced to their rank-2 irreducible tensor contribution. We have indeed observed a strong interplay between the rank-2 tensor and the central terms originating from the Lorentz PV and T interactions. While these two terms generally appear separately in non-relativistic nuclear models, they originate from the same interaction vertex in the relativistic ones. In the intermediate mass nuclei that we have explored, the s.p. energies are shown to be very much impacted by the Lorentz PV and T interactions which impose some very specific behavior along isotonic chains. Even their weak influence results from the cancellation of two of their constituent terms. The effect of the Lorentz PV and T couplings --- which are purely relativistic terms --- in s.p. energies is therefore very different from the rank-2 tensor force.
We conclude from this study that, while both relativistic and non-relativistic models for the nuclear interaction could equally well reproduce global properties of finite nuclei, such as binding energies and radii, the detailed evolution of s.p. energies could potentially sign relativistic effects such as the Lorentz PV and T interactions. This statement requires however further analysis based on other modelings of the nuclear interaction, but we propose here a clear relation between Lorentz PV and T relativistic vertices and s.p. evolution.

%
\section*{Acknowledgements}
J. L. acknowledges the support by China Scholarship Council (CSC).
This work has been partially funded by the NewCompStar COST action MP1304, the SN2NS project (ANR-10-BLAN-0503),
and by the National Natural Science Foundation of China (Grant Nos. 11375076 and 11405223).


\end{document}